\def\lessapprox{\,\raise 0.6ex\hbox{$<$}\kern -0.75em\lower 0.47ex
    \hbox{$\sim$}\,}
\def\largapprox{\,\raise 0.6ex\hbox{$>$}\kern -0.75em\lower 0.47ex
    \hbox{$\sim$}\,}
\def\ni{\noindent}
\def\wmm{\omega_{\mu\mu}(\theta)}
\def\wmk{\omega_{\mu\kappa}(\theta)}

\def\cpp{C_{pp}(\theta)}
\def\ckk{C_{\kappa\kappa}(\theta)}

\def\be{\begin{equation}}
\def\ee{\end{equation}}
\def\bea{\begin{eqnarray}}
\def\eea{\end{eqnarray}}

\def\eps2{{\epsilon^2}}

\def\eg{{\sl eg.\ }}
\def\ie{{\sl i.e.\ }}
\def\etal{{\sl et al.\ }}
\def\b91{Blandford \etal (1991)}

\def\xp{x^{\prime}}

\def\vx{{\underline x}}

\def\go{\mathrel{\raise.3ex\hbox{$>$}\mkern-14mu
             \lower0.6ex\hbox{$\sim$}}}
\def\lo{\mathrel{\raise.3ex\hbox{$<$}\mkern-14mu
             \lower0.6ex\hbox{$\sim$}}}

\def\w{\omega(\theta)}

\def\wkk{\omega_{\kappa\kappa}(\theta)}
\documentstyle[onecolumn]{mn}
\begin{document}
\title[Clustering from Weak Gravitational Lensing]
{Clustering of Faint Galaxies: $\w $,\\
Induced by Weak Gravitational Lensing}
\author[Jens Verner Villumsen]
{Jens Verner Villumsen\\
Max Planck Institut f\"{u}r Astrophysik\\
Karl Schwarzschild-Str. 1, Postfach 15 23,\\
85740 Garching, Germany\\
jens@mpa-garching.mpg.de}
\maketitle
\begin{abstract}
Weak gravitational lensing by large scale structure affects the
number counts of faint galaxies through
the ``magnification bias'' and thus affects the measurement of the
angular two-point correlation function $\w $.
At faint magnitudes the clustering amplitude will
decrease differently with limiting magnitude than expected from Limber's
equation.  The amplitude will hit a minimum and then rise with
limiting magnitude.  This
behavior occurs because $\w$ due to clustering decreases with distance,
while the ``magnification bias'' due to weak lensing increases
with distance. The apparent magnitude $m_{min}$ at which the
magnification bias starts to dominate the observed clustering
 is model and color dependent.  It is given by
$\omega(m=m_{min},\theta=5^\prime) \approx (1\ -\ 2)\times
10^{-3}(5s-2)^2 \Omega_0^2 \sigma_8^2$, where  $s$ is the logarithmic
slope of the number counts.  Already published measurements of $\w$ at
$R=25$ may be strongly influenced by the ``magnification bias''.
An experiment using the ratio of blue and red number counts across the
sky can be designed such that the effects of the ``true'' clustering
is minimized. The magnification bias is a measurement
of the clustering of the mass.  This weak lensing experiment does not
require measuring shapes and position angles of galaxies.  I derive a
revised Limber's Equation including the effects of magnification bias.
\end{abstract}

\noindent Submitted to: Monthly Notices of the Royal Astronomical Society

\section{Introduction}

One of the most useful statistics for the evolution and
clustering properties of faint, distant galaxies is the two-point
angular correlation function $\w $.  It measures the excess
number of pairs of galaxies separated by angle $\theta$ on the sky.
For a given galaxy sample, $\w $ depends on the redshift
distribution of the galaxies, and the three dimensional clustering
amplitude $\xi(r,z)$.  As pointed out by Koo \& Szalay (1984),
$\w $ can be used to constrain models of the evolution and
clustering of faint galaxies at intermediate and high redshift.  A
large number of investigations have measured the correlation for faint
magnitudes in different passbands, \eg Bernstein \etal 1994; Brainerd,
Smail, \& Mould 1995 (BSM); Couch, Jurcevic, \& Boyle 1993;
Efstathiou \etal 1991; Neuschaefer, Windhorst, \& Dressler 1991;
Roche \etal 1993.  There is a general consensus that $\w $
can be approximated by a power law with slope of $-0.8$ and an
amplitude that decreases with increasing limiting magnitude.  There
is, however, little consensus on what that means in terms of
luminosity evolution, merger history, and clustering.  The number
counts as a function of magnitude, (\eg Lilly \etal 1991, Smail \etal
1995), and the redshift
surveys to faint magnitudes (\eg Broadhurst \etal 1992, Glazebrook
\etal 1994) put significant constraints on the galaxy evolution.

Weak gravitational lensing of distant galaxies, equivalent to a
systematic induced magnification and distortion without multiple
imaging, is a probe of the large scale mass distribution of the
universe, (\eg Gunn 1967).  Essentially, when a light
ray from a distant galaxy passes through an overdense region in mass,
the image will be magnified and distorted tangentially with respect to
the center of the mass overdensity.  This effect has been measured in
a number of galaxy clusters, [\eg Tyson, Valdes, \& Wenk (1990),
Bonnet \etal (1994)].  Conversely, passing
through a low density region, the image will become dimmer and
elongated radially. Independent perturbations along the line of sight
add up stochastically [Blandford \etal 1991(BSBV), Miralda-Escud\'{e} 1991,
Kaiser 1992, Villumsen 1995a(V95a)].  Weak lensing by large
scale structure is observable because galaxy images near each other on the sky
will be sheared coherently leading to a locally preferred direction.
Due to the weak
nature of the lensing there has been no undisputed measure of
the distortion, see f.ex. (Mould \etal 1994; Kaiser, Squires, \&
Broadhurst 1995),
though a reanalysis by Villumsen (1995b) shows a tentative detection at
the 5-$\sigma$ level consistent with expectations from a weakly biased
flat Cold Dark Matter (CDM) universe.
The problem with weak lensing is that it is
weak.  Measuring the weak shear requires measuring the shape and
orientation of distant galaxy images.  This is observationally
feasible, but quite difficult,
since the images are small and faint, and the possible systematic
errors are the limiting factors.

Weak gravitational lensing is characterized by the convergence
$\kappa$ and shear $\gamma$.  We will work only
in the weak limit, \ie $|\kappa|,|\gamma| \ll 1$.  In this limit, both
$\kappa$ and $\gamma$ are observables.  The distortion of an image is
given by $\gamma$, such that an intrinsically round image will aquire
an eccentricity equal to $|\gamma|$ and the phase of $\gamma$ will be
twice the position angle of the major axis.  The magnification $A$ of
an image is
$A=1+2\kappa$.  In clusters, the distortion is used to generate shear
maps which can then be converted into maps of the
surface density as pioneered by Kaiser \& Squires (1993) and further
developed by Seitz \& Schneider (1995).  Recently Broadhurst, Taylor \&
Peacock(1995) have  demonstrated the use of the magnification to improve the
information about the surface mass density.

In this paper I propose a new  measure of weak gravitational lensing by
large scale structure based on the magnification bias introduced by
Turner (1980) and used by Broadhurst (1995) for lensing in
clusters.  Basically, weak lensing will induce angular correlations in
the number density of galaxies on the sky.
This measure only involves the convergence $\kappa$ and does not
involve $\gamma$.  Thus, it is not necessary to measure the shape of
high-redshift galaxies.  This means that a galaxy survey to look for
magnification bias can be pushed to fainter magnitudes than a search
for galaxy distortions.

In \S2 I estimate the amplitudes of ``true'' and ``apparent''
clustering, in \S3 I work out the theory.  In \S4 I calculate $\w$ in
terms of the 3-D correlation function and in \S5 I discuss the
implications.

\section{Estimates of ``true'' and ``apparent'' clustering}

\bigskip

Magnification bias is the change in surface number density of sources on the
sky due to gravitational lensing (\eg Turner \etal 1984).  As the total
number of galaxies on
the sky is conserved, magnification in a given patch of the sky will
lead to a decrease in the number density of galaxies, because you are
looking at a smaller area of sky than you think.  However,
magnification will also brighten an image so more galaxies will
be visible at a given apparent magnitude $m$.  If the number counts
$N_0(m)$ in the absence of lensing has a slope
\be
s={d\; \log N_0(m) \over dm},
\ee
then the number counts will be changed
\be
N_{obs}(m)=N_0(m)A^{2.5s-1}=N_0(m)(1+(5s-2)\kappa),\label{Nobs}
\ee
where $N_{obs}(m)$ and $N_0(m)$ are the number counts in the
presence/absence of lensing.  The first equality is generally valid,
the second assumes that $|\gamma|,|\kappa|\ll 1$.  For
$s=0.4$ there is no magnification bias,
for $s\ <\ 0.4$ there is a depletion of counts, while for $s\ >\ 0.4$
there is an increase in counts in the presence of magnification.

In order to estimate the clustering effects of the ``magnification bias'',
assume a universe with no
intrinsic clustering of galaxies.  Then weak lensing will introduce an
apparent clustering characterized by an angular two-point correlation
function $\wkk$
\bea
\wkk &\equiv&{\left<\left[N_{obs}(\bar\phi+\bar\theta,m)-N_0(m)\right]
\left[N_{obs}(\bar\phi,m)-N_0(m)\right]\right> \over N_0(m)^2}.\label{wkk}
\eea
Here the correlation function is calculated as the average over all
direction vectors $\bar\phi$ of the relative excess number of galaxy
pairs separated by angle $|\bar\theta|=\theta$.  We obtain from
Eqs. (\ref{Nobs},\ref{wkk})
\bea
\wkk &=& {\left<\left[N_0(m)(5s-2)\kappa(\bar\phi+\bar\theta)\right]
\left[N_0(m)(5s-2)\kappa(\bar\phi)\right]\right> \over N_0(m)^2}\\
&=&(5s-2)^2\left<\kappa(\bar\phi+\bar\theta)\kappa(\bar\phi)\right>
\equiv (5s-2)^2 C_{\kappa\kappa}(\theta)\;=\;(5s-2)^2 C_{pp}(\theta).
\eea
Here, $\ckk$, and $\cpp$ are the correlation functions
of the convergence and the shear.  In this weak limit they are
identical (BSBV and V95a).   The shear and convergence correlation
functions are
steeply increasing functions of the comoving angular distance $y$, and
thus of redshift $z$. For a universe with $\Omega_0=1,\;\Omega_\Lambda=0$,
\be
\omega_{\kappa\kappa}\propto C_{\kappa\kappa}\propto y^3 =
8\left(1-(1+z)^{-1/2}\right)^3.
\ee
The variation in the number counts to a given magnitude limit across
the sky is an observable.  The shear and convergence correlation
functions are related to the density distribution and cosmological
parameters as described in BSBV, V95a.  The shear $\gamma$ will also
influence the number counts, but the effect is quadratic in $\gamma$ and
can be ignored.

In real life, of course, we observe a true clustering of galaxies at
faint magnitudes characterized by the angular two-point correlation function
$\wmm$, (see references in \S1).  The correlation induced by
magnification bias $\wkk$ is of practical interest only  if it becomes
significant compared to $\wmm$. In a universe
where the clustering is constant in comoving coordinates, $\wmm$
will fall off roughly inversely with distance.  Lensing is a cumulative effect
while the intrinsic three dimensional clustering will be
diluted by projection effects.  This would indicate that at low
redshift $\wkk$ is unimportant but that at sufficiently high redshift
$\wkk$ might become important.  The observational issue is whether
this occurs at observable apparent magnitudes.

The current status of $\w $ for red samples is summarized by (BSM),
see Fig. 2.  It is common to characterize the correlation function as a
power law with slope
$\gamma$ and the amplitude at 1 degree, $A_w$.  Then $\w \approx A_w
\theta^{-\gamma}$, $\theta$ measured in degrees.  There is a general
consensus that $\gamma\approx
0.8$ and that $A_w$ is a decreasing function of depth.  (BSM)
state that the amplitude from red counts is
$A_w \propto R^{-0.27 \pm 0.01}$ with $A_w(R=25)\approx 3\times
10^{-4}$.

For a CDM universe where the characteristic redshift of the sources is $z\
\sim\ 1$, the amplitude of the polarization correlation is given by
the cosmological density parameter $\Omega_0$ and the rms density
fluctuations on a scale of 800 km $s^{-1}$, $\sigma_8$ (BSBV, V95a),
\be
C_{\kappa\kappa}(\theta=5^\prime)\approx 1.5\times 10^{-3}\Omega_0^2
\sigma_8^2.
\ee
This would indicate that the amplitude of $A_w$ induced by lensing is
\be
A_{w,\kappa}\approx 2\times 10^{-4} (5s-2)^2\Omega_0^2 \sigma_8^2 .
\ee
The Mould \etal (1994), presented in (BSM) were
obtained in the $r$ band where the source counts have $s_r=0.3\ $.  Then
\be
A_{w,\kappa}(r)\approx 5\times 10^{-5}\Omega_0^2\sigma_8^2.
\ee
This is lower than the observed value of the correlation function
$A_w(R=25)\approx 3\; 10^{-4}$, $A_w(R=25.5)\approx 1.5\; 10^{-4}$,
but not negligible.
However, we have neglected the correlation between the
magnification bias and the true clustering.  If light traces mass,
then there will be a strong correlation between the magnification and
the true clustering.  If $s<0.4$, this is really an anticorrelation
and the observed clustering will be less than the true clustering.

Smail \etal (1995) find that the slope in the VRI bands all tend to
$s \approx 0.3$ for the faintest galaxies.
Neuschaefer \etal (1991) have reported on $\w$ in the Gunn-g band and they find
a slope $s_g\approx 0.45$.  This is close to the critical slope
of $s=0.4$, so the magnification bias is expected to be small and
quite uncertain.  For $g<24.5$ they find that $\w$ is approximately a power
law with the amplitude decreasing with increasing magnitude.  For
fainter magnitudes they find a very steep rise in the amplitude.  This
rise is much too steep to be accounted for by magnification bias.  At
$g\approx 24.5$, $\omega(\theta=1')\approx 10^{-2}$.  For a high value
of $\Omega_0\;\sigma_8$ there may be a small contribution to $\w$ from
magnification bias, though this is uncertain.

However, Broadhurst (1995) has shown that for the $I$ band counts, $I\ >\
24$, the red counts, $V-I\ >\ 2.0$, are fitted well by a slope
$s_r \approx 0.15$.  The blue counts, $V-I\ <\ 1.0$, have $s_b \approx 0.5\ $.
The prediction is thus that for the blue counts, there will be a measurable
magnification bias.  For the red counts, though, the prediction is
that the magnification bias will be strong,
\be
A_{w,\kappa}(r)\approx 3\times 10^{-4}\Omega_0^2\sigma_8^2.
\ee
This is comparable to the observed correlation function amplitude.

These results indicate that at accessible apparent magnitudes the
magnification bias will give a significant contribution to the observed
angular two-point correlation function.  This encourages a more detailed
study of the magnification bias.

\section{Theory}

\subsection{Single Sample Statistics}

In the limit
where the intrinsic clustering and the magnification bias are weak, the
clustering, characterized by the angular two-point correlation
function $\w $ can be calculated.  The magnification bias works
on the ``true'' number counts, \ie the number counts which include
the true clustering,
\be
N_{obs}(m)=N_0(m)(1+\Delta\mu)A^{2.5s-1}\approx
N_0(m)(1+\Delta\mu+(5s-2)\kappa) .
\ee
Here, $\Delta\mu$ is the true fractional excess number of sources in a
particular patch on the sky.  From this we can calculate $\w $ from an average
over the sky,
\bea
\w  &=&{\left<\left[N_{obs}(\bar\phi+\bar\theta,m)-N_0(m)\right]
\left[N_{obs}(\bar\phi,m)-N_0(m)\right]\right> \over N_0(m)^2}\\
&=&\left<\left[\Delta\mu(\bar\phi+\bar\theta)+
(5s-2)\kappa(\bar\phi+\bar\theta)\right]
\left[\Delta\mu(\bar\phi)+(5s-2)\kappa(\bar\phi)\right]\right>\\
&=&\left<\Delta\mu(\bar\phi+\bar\theta)\Delta\mu(\bar\phi)\right>
+(10s-4)\left<\Delta\mu(\bar\phi+\bar\theta)\kappa(\bar\phi)\right>
+(5s-2)^2\left<\kappa(\bar\phi+\bar\theta)\kappa(\bar\phi)\right>\\
&\equiv&\wmm+2\wmk+\wkk\\
&=&\wmm+2\wmk+(5s-2)^2 C_{\kappa\kappa}(\theta).
\eea
Again, the average is over all direction vectors $\bar\phi$.  The
three individual terms are not observables, only the sum.  The first
term is the true galaxy clustering, the third term is the apparent
clustering induced by the magnification bias, and the second term is a
cross correlation.  The magnification bias is induced by mass
density fluctuation, while the true galaxy clustering measures the
galaxy number density fluctuations.  If in projection, the number
density fluctuations are decoupled from the mass density fluctuation,
then the cross term will be zero.

The observed number counts of galaxies as a function of magnitude in a
given passband and position on the sky will depend in a complicated,
and unknown way, on the formation and evolution of galaxies.
In the calculations we are going to use the comoving radial distance
$x$ as fundamental variable,
\bea
x(z)&=&\int_0^z dz'\; H^{-1}\left(z'\right)\ ;\
H(z')=\left[\Omega_0(1+z')^3+(1-\Omega_0-\Omega_\Lambda)(1+z')^2+
\Omega_\Lambda\right]^{1/2},\\
y(z)&=&{\sinh\left[\sqrt{(1-\Omega_0-\Omega_\Lambda)}\; x(z)\right] \over
\sqrt{(1-\Omega_0-\Omega_\Lambda)}}.
\eea
In a
simple model, characterised by a selection function $S(x)$, and linear
bias factor $b(x)$ we can calculate statistically the surface density
distribution.  We define $S(x)$ to be the mean comoving number density
of observed objects at distance $x$.  The function $S$ can also be
thought of as a function of redshift, or lookback time.
The normalisation of $S$ is such that
\be
\int_0^{\infty}dx\;y^2 S(x)=1. \label{normS}
\ee
The function $S$ has hidden in it the
number density evolution, the luminosity evolution, and the
differential K-correction.  We define the linear bias factor $b$ as
\be
b(x)=\left({\delta N\over N}\right)\bigg/\left({\delta\rho \over \rho}\right),
\ee
so $b$ is not a function of scale but can be a function of epoch.
This linear biasing scheme assumes that the galaxies cluster the same
way as the mass, just at a possibly different amplitude characterized by $b$.

The surface density fluctuation $\Delta\mu$
can then be written as a line integral in comoving radial distance $x$
of the fractional volume overdensity
$b(x)\delta(\vx)$ times the volume element $dx\cdot y^2$ with a selection
function $S(x)$.  Here $y$ is the comoving angular diameter distance.
\bea
\Delta\mu(\bar\theta) &=& \int_0^{\infty} dx\;  y^2
S(x)b(x)\delta(\vx), \\
\kappa(\bar\theta)&=&3\Omega_0\int_0^{\infty}dx\;  y
w(x)\delta(\vx)/a.
\eea
In principle, all the line integrals are only out the horizon distance
$x_H$, however, we can just set $S(x)$ equal to zero for $x > x_H$.
The equation for $\kappa$ is taken from V95a, Eq 29, where
$\kappa \equiv -\Delta M$.  Here, $\delta(\vx)$ is the density
perturbation at position $\vx$ evaluated at the epoch when the
lightray passes through.  The expansion factor $a$ also enters, $a=1$
at the present epoch.  The function $w(x)$ is the lensing
selection function which for a lens at distance $x$ is the integral over all
sources $y^2 S(\xp)$ further away than $x$ of the ratio of the
lens-source distance
$y_{LS}$ and the observer-source distance $y_{OS}$.
\be
w(x')=\int_{x'}^{\infty}dx\; y^2 S(x){y_{LS} \over y_{OS}}
\ee
For convenience we write
$\delta(\vx)\equiv f(x)a \delta_0(\vx)$, where $\delta_0(\vx)$ is the
density fluctuation evaluated today.  We implicitly assume that the
``growth factor'' $f(x)$ is a universal function of expansion factor $a(x)$
only.  Then
\bea
\Delta\mu(\bar\phi)+(5s-2)\kappa(\bar\phi)=\int_0^{\infty}dx\;y\;f(x)
\left[y\;S(x)a\;b(x)+3\Omega_0(5s-2)w(x)\right]\delta_0(\vx). \label{deltam}
\eea

Suppose we have sufficiently well behaved functions $F(\bar\phi)$, and
$G(x)$  defined as
\be
F(\bar\phi)\equiv \int_0^{X_H}dx G(x) \delta_0(\vx).
\ee
The two-point correlation function $C_{FF}(\theta)$ of $F$ will then be
\be
C_{FF}(\theta)=4\pi^2\int_0^{\infty}dx\; G^2(x)\int_0^\infty dk\;k\;P(k)
J_0(k x \theta). \label{FF}
\ee
Equation \ref{deltam} is of this form so we can write the two-point
correlation function as
\bea
\w &=&4\pi^2\int_0^{\infty} dx\ y^2\ f^2(x)\left[ y S(x)
a\;b(x)+3\Omega_0(5s-2)w(x)\right]^2
\;\int_0^\infty dk k P(k) J_0(kx\theta). \label{w1band}
\eea

We see in Eq.(\ref{w1band}) that if $s>0.4$, the weak lensing will
give a positive contribution to the correlation amplitude, while if
$s<0.4$, the weak lensing will have a negative contribution to
the correlation amplitude.

The weak lensing can change the angular dependence of $\w$ if $P(k)$
is not a power law.  On a given angular scale $\theta$, $\w$ samples
decreasing wavenumbers as a function of $x$ as seen in
Eq.\ref{w1band}.  The weighting in $k$ is given by the outer integral
which is changed by the magnification bias.  Thus, if $P(k)$ is not a
power law $\w$ will be changed by the magnification bias.

At low redshift the intrinsic clustering will dominate and the
magnification bias is negligible.  At higher redshift, the intrinsic
clustering will drop and the magnification bias will increase
dramatically.  In the case of flat number counts, \ie $s<0.4$, the angular
two-point correlation function will decrease faster with depth than
expected from Limber's equation (Peebles 1980).
Eventually the contibutions from the true clustering and the
magnification will be about equal and the observed clustering
amplitude will hit a minimum.  At even higher redshift the
magnification bias will dominate and $\w $ will increase.

For steep number
counts, \ie $s>0.4$, the correlation amplitude will fall slower
with apparent magnitude than expected from Limber's equation, reach a
minimum, and then rise again.  In both cases the correlation
amplitude will eventually become an increasing function of depth.  The
magnitude at which this happens will depend on cosmological
parameters, $\Omega_0$ and $b$.

How do the intrinsic correlations and the apparent correlations scale
with depth of the sample.  Take a simple example where $\Omega_0=1$,
$\Omega_\Lambda=0$, $1-a \ll 1$, and $b(x)$ is a constant, then $y=x$.
In Eq.(\ref{w1band}) in the outer integral, the contribution
from intrinsic clustering will scale inversely with $q$, while the contribution
from magnification bias will scale as $q^3$.  In Eq. \ref{w1band} the
calculation of the intrinsic clustering involves five powers of the
distance.  As $S^2$ is inversely proportional to the sixth power of
the characteristic distance of the sources we have that $\wmm$ scales
inversely with characteristic distance of the sources.  Also in
Eq. \ref{w1band} we see that $\wkk$ scales as the third power of the
characteristic distance.  The inner integral in $k$ is the same for
the two terms, and is proportional to $x^{-2-n}$ for a power law power
spectrum with slope $n$.

We can thus estimate the
sample distance at which the magnification bias will dominate.
The correlations induced by weak lensing are approximately
$\wkk \propto \Omega_0^2 \sigma_8^2 y^{1-n}$ while the true angular
correlation has $\wmm \propto \sigma_8^2 b^2
y^{-3-n}$.  Thus the characteristic distance $y_e$ at which the two
effects become comparable is $y_e^{-2}\propto \Omega_0/b$.  The
limiting magnitude $m_{min}$ at which the correlation amplitude is at
a minimum is an observable.  If the typical sample redshift is
$z\approx 1$ then $m_{\rm min}$ is determined from the equation.
\be
\omega(m=m_{\rm min},\theta=5^\prime) \approx (1\ -\ 2)\times
10^{-3}(5s-2)^2 \Omega_0^2 \sigma_8^2.
\ee

In Figure 1 we demonstrate the effects of the magnification bias
through a simple model.  Assume that $S(x)$ is
constant out to some distance $x_0$.  Assume that $b=1/\sigma_8$, \ie
the bias is independent of epoch.  Further assume $\Omega_0=1$ and
$P(k)\propto k^{-1.2}$.  In this model, which should be seen as an
illustration only, we can calculate $A_W$ as a function of mean source
redshift $<z>=\int_0^\infty dx\;z(x) x^2 S(x)$, for various values of $s$
and $b$.  In all these models the intrinsic clustering is the same.
\input{psfig}
\begin{figure}
\psfig{figure=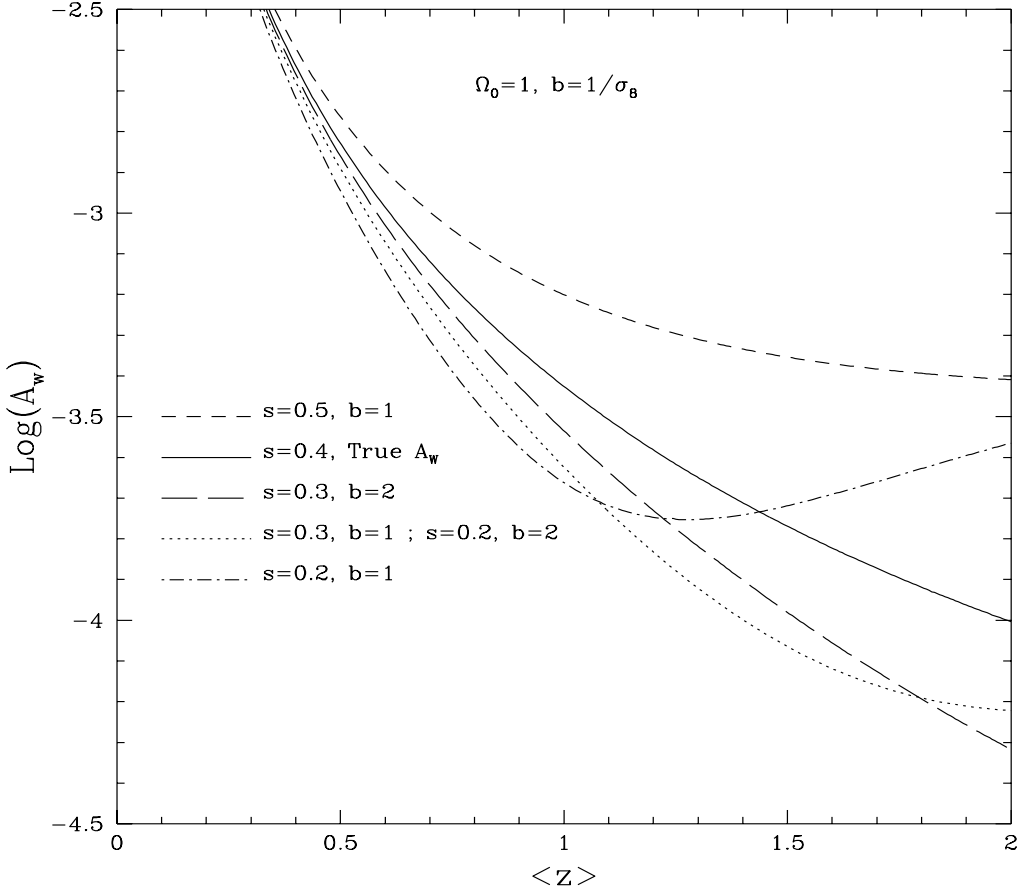,height=14cm,width=14cm}
\caption{Correlation Amplitude $A_w$ as a function of mean redshift
 and slope of number counts.  The solid curve is the intrinsic
correlation.  The dotted, long dashes, dash-dotted curves are relevant
for red galaxy samples.  The short-dash curve is relevant for blue
galaxy samples.}
\end{figure}
The solid curve, $s=0.4$, is due entirely to the intrinsic clustering,
there is no magnification bias.  The uppermost curve, $s=0.5,\;b=1$,
shows a positive magnification bias and falls off slower than the true
clustering.  At $<z>\approx 1$ the observed clustering amplitude is
70\% higher than the true amplitude.  At higher redshifts $A_w$ is
quite flat and at $<z>\approx 2.5$ it will begin to rise.

The two
curves for $s=0.3$, which are appropriate for red samples, originally
fall faster than the true correlation.  For $z \lessapprox 1.8$, the
less biased model, dotted line, will have a lower amplitude of $A_w$
than the more biased model, while at higher redshifts the opposite is true.
This behavior can be understood as follows.  The magnification bias
consists of two terms, $\wkk$ which is always a positive contribution
and quadratic in $\sigma_8$, and $\wmk$ which is linear in $\sigma_8$
and is negative.  At low redshift the negative linear term dominates
while at sufficiently high redshift the positive quadratic term
dominates.

For an unbiased population of galaxies with $s=0.2$, the magnification
bias is so strong that the correlation amplitude hits a minimum at
$<z>\approx 1.3$ and is larger than the true amplitude for
$<z> \largapprox 1.5$.  The curve for $b=2$, $s=0.2$ is identical to the
curve for $b=1$, $s=0.3$.

If we observe the correlation function $\w$ for a single limiting
magnitude we cannot tell what is the intrinsic contribution and
what is magnification bias, it is necessary to use a range of
limiting magnitudes.  The slope of the number counts $s$ is an
observable, so if $s < 0.4$, and the correlation amplitude is
decreasing with limiting magnitude, we know that the intrinsic
correlation is higher than the observed correlation.  If the amplitude
is rising we cannot make this inference.  If $s > 0.4$ we know that
the true correlation is less than the observed correlation.

If an increase in correlation amplitude as a function of limiting
magnitude is observed, we see that it is not necessary to invoke a
local population of weakly clustered intrinsically faint galaxies.  It
is also not necessary to assume that we are seeing a strongly
clustered population of distant galaxies.

These models should be seen as an illustration only.  In order to
properly model the correlation function we need to consider luminosity
evolution, merger history, density evolution, differential
K-correction etc.  In particular we need to relate $<z>$ to limiting
magnitude.

{}From Figure 1, we see that if the data at $R \approx 25.0-25.5$ in
(BSM) is at $<z> \approx 1$, there is likely to be a
significant contribution from magnification bias in the data.

\subsection{Two-Sample Statistics}

An equivalent statistic involving two samples can be used.  The number
counts of sources selected in two separate ways in the sample field
are used.
The ratio of counts in two passbands at magnitudes $(m_1,m_2)$ can be
calculated in the case of weak intrinsic clustering of galaxies.
\bea
\Delta(m_1,m_2,\bar\phi)&\equiv& {N_{obs}(m_1,\bar\phi)\over
N_0(m_1)}\bigg/ {N_{obs}(m_2,\bar\phi) \over N_0(m_2)}-1
\approx{1+\Delta\mu(m_1,\bar\phi)+[5s_1-2]\kappa_1(\bar\phi)\over
1+\Delta\mu(m_2,\bar\phi)+[5s_2-2]\kappa_2(\bar\phi)}-1 \\
&\approx&\Delta\mu(m_1,\bar\phi)-\Delta\mu(m_2,\bar\phi)
+5\left[s_1\kappa_1(\bar\phi)-s_2\kappa_2(\bar\phi)\right]
-2\left[\kappa_1(\bar\phi)-\kappa_2(\bar\phi)\right].
\eea
$(m_1,m_2)$, $(s_1,s_2)$, $(\kappa_1,\kappa_2)$ are the magnitudes,
slopes of the number counts and the convergences in the two passbands.  The
convergences need not be identical since the depth of the samples in
redshift need not be the same in the two passbands.  Equivalently, we
can look at two separate samples selected by color, in a single passband.
This statistic in the weak clustering regime is equivalent to the
statistic involving the difference in number counts.
\bea
\Delta(m_1,m_2,\bar\phi)&\equiv& {N_{obs}(m_1,\bar\phi)\over
N_0(m_1)} - {N_{obs}(m_2,\bar\phi) \over N_0(m_2)}\\
&=&\Delta\mu(m_1,\bar\phi)-\Delta\mu(m_2,\bar\phi)
+5\left[s_1\kappa_1(\bar\phi)-s_2\kappa_2(\bar\phi)\right]
-2\left[\kappa_1(\bar\phi)-\kappa_2(\bar\phi)\right].
\eea

The correlation function can then be calculated in a similar fashion
to Eq. (\ref{w1band})
\bea
\w &=&4\pi^2\int_0^{\infty}dx\;y^2 f^2(x)
\left[y\;a\left(S_1(x)b_1(x)-S_2(x)b_2(x)\right)
+3\Omega_0\left(5(s_1 w_1(x)-s_2 w_2(x))
-2\left(w_1(x)-w_2(x)\right)\right)\right]^2 \nonumber \\
& &\times\int_0^\infty dk\;k\;P(k) J_0(kx\theta). \label{w2band}
\eea

\begin{figure}
\psfig{figure=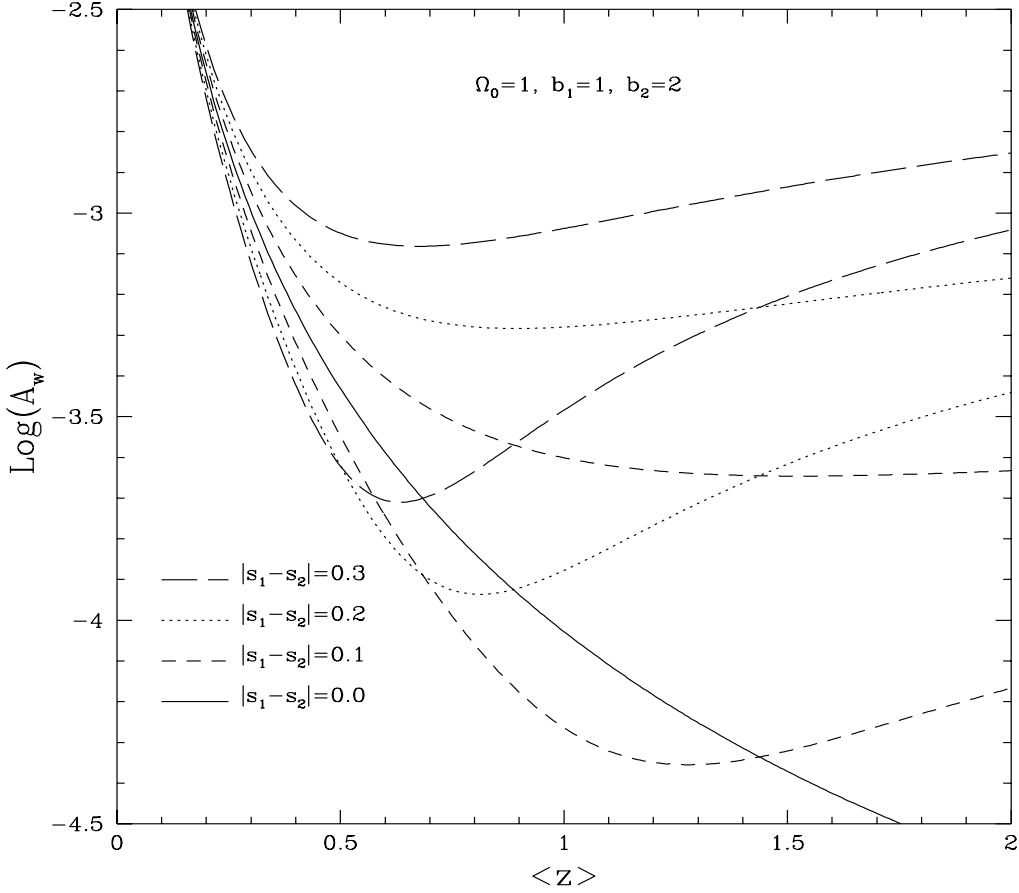,height=14cm,width=14cm}
\caption{Correlation Amplitude $A_w$ as a function of mean redshift
 and difference in slope of number counts.  The three curves that
start below the solid curve are for $s_1<s_2$.  The three other curves
are for $s_1>s_2$.}
\end{figure}

Figure 2 shows the correlation amplitude for the same model model as
in Figure 1.  Here we have assumed two populations with the same
selection function $S_1(x)=S_2(x)$, thus $w_1(x)=w_2(x)$.  We have
further assumed that $\sigma_8=1$ and that $b_1=1$, $b_2=1/2$.  This
means we have strongly and weakly correlated populations.  We show
$A_w$ for $\Delta s=s_1-s_2=(-0.3,-0.2,-0.1,0,0.1,0.2,0.3)$.  The solid curve
is $A_w$ in the absence of magnification bias $\Delta s=0$.  For a
positive value of $\Delta s$ the curves are always above the solid
curve.  At low redshift there is little magnification bias and the
curves all follow the solid curve.  For positive $\Delta s$ the curves
move significantly away from the solid line at $<z> \approx 1/2$ and
then become nearly flat out to $<z> \approx 2$.  For negative $\Delta
s$, the curves trace the solid line more closely, but with a
significant difference at $<z> \approx 1/2$.  For the curves with a
substantial difference in number counts slope, the curves then hit a
minimum and then rise steeply.

These curves for negative $\Delta s$ may be relevant for a strongly
clustered red population with small number counts slope and a weakly
clustered blue population with a larger value of $s$.

This estimator is most useful when the intrinsic clustering is similar
in the two bands, while the slopes of the number counts are very
different.  In the case
where the intrinsic clustering is the same in the two passband, \ie
$\Delta\mu(m_1,\bar\phi)=\Delta\mu(m_2,\bar\phi)$, there is
zero contribution from the intrinsic clustering. If also the depth in
comoving distance is similar, then
$\kappa_1\approx \kappa_2$ and we get that
\be
\wkk=25(s_b-s_r)^2 \ckk . \label{wratio1}
\ee
This assumption is not as severe as it might
seem as can be seen from the following example.  Look at two source
planes at distances $x_1 < x_2$ and two lightrays separated by a small
angle $\bar\theta$.  A mass fluctuation at distance $x$ will
coherently influence the convergence in the two beams only if
the beams both pass through the density fluctuation.  This is
obviously only possible if $x < x_1$.  Thus density fluctuations beyond
the nearest source plane will not influence the correlation, and
\be
\left<\kappa_1(\bar\phi+\bar\theta)\kappa_2(\bar\theta)\right>\approx
\left<\kappa_1(\bar\phi+\bar\theta)\kappa_1(\bar\theta)\right>
\Rightarrow
C_{\kappa_1\kappa_2}(\theta) \approx C_{\kappa_1\kappa_1}(\theta).
\ee
In that case we have to a good approximation that the observed two
point correlation function of the ratio of counts is given by
Eq. \ref{wratio1}, which for the ratio of very red and very blue
samples gives
\be
\wkk \approx 3 \ckk. \label{wratio2}
\ee

If at intermediate magnitudes we find that the observed correlation
strengths differ by a factor $C^2$ we can use a different estimator of
the clustering,
\bea
\Delta(m_1,m_2,\bar\phi)&\equiv& \left[{N_{obs}(m_1,\bar\phi)\over
N_0(m_1)}-1\right]\ -C\left[{N_{obs}(m_2,\bar\phi) \over N_0(m_2)}-1\right]\\
&=&\Delta\mu(m_1,\bar\phi)-C\Delta\mu(m_2,\bar\phi)
+5\left[s_1\kappa_1(\bar\phi)-C\;s_2\kappa_2(\bar\phi)\right]
-2\left[\kappa_1(\bar\phi)-C\;\kappa_2(\bar\phi)\right].
\eea
The correlation function can then be calculated in a similar fashion
to Eq. (\ref{w1band})
\bea
\w &=&4\pi^2\int_0^{\infty}dx\;y^2 f^2(x)
\left[y\;a\left(S_1(x)b_1(x)-C\;S_2(x)b_2(x)\right)
+3\Omega_0\left(5(s_1 w_1(x)-C\;s_2 w_2(x))
-2\left(w_1(x)-C\;w_2(x)\right)\right)\right]^2 \nonumber \\
& &\times\int_0^\infty dk\;k\;P(k) J_0(kx\theta). \label{w2Cband}
\eea
By construction, the intrinsic correlations cancel out at intermediate
redshift.  At higher redshifts this cancellation may not occur, but
this estimator may minimize the effects of the intrinsic clustering
even at faint magnitudes.

\section{Correlation Functions, Revised Limber's Equation}

Suppose we have sufficiently well behaved functions $F_1(\bar\phi)$,
$F_2(\bar\phi)$, $\delta_1(\vx)$, $\delta_2(\vx)$, $G_1(x)$, and
$G_2(x)$ defined as
\bea
F_1(\bar\phi)&\equiv& \int_0^\infty dx G_1(x) \delta_1(\vx)\ ;\
F_2(\bar\phi)\equiv \int_0^\infty dx G_2(x) \delta_2(\vx)\ ;\
F(\bar\phi)\equiv F_1(\bar\phi)+F_2(\bar\phi).
\eea
The two-point correlation function $C_{FF}(\theta)$ of $F$ will then be
\bea
C_{FF}(\theta)=2\int_0^\infty dx& &\left(G_1^2(x)\int_0^\infty
dq\; \xi^{11}\left(\left((x\theta)^2+q^2\right)^{1/2}\right]\right)+\nonumber\\
& &\left(G_2^2(x)\int_0^\infty
dq\;
\xi^{22}\left[\left((x\theta)^2+q^2\right)^{1/2}\right]\right)+\nonumber
\\
& &\left( 2G_1(x)G_2(x)\int_0^\infty
dq\; \xi^{12}\left[\left((x\theta)^2+q^2\right)^{1/2}\right]\right),
\eea
where $\xi^{11}$, $\xi^{22}$, and $\xi^{12}$ are the two
autocorrelation functions of $\delta_1(\vx)$ and $\delta_2(\vx)$ and
their cross correlation function.  This follows from Eq. \ref{FF} and
V95a, Eqs. 37-39.

Let us take some galaxy density distribution $\delta^g(\vx)$ and mass
density distribution $\delta^m(\vx)$.  Then the surface density of
galaxies $\Delta\mu$ and convergence $\kappa$ are
\bea
\Delta\mu(\bar\theta) &=& \int_0^{\infty} dx\;  y^2
S(x)\delta^g(\vx),\label{dmu} \\
\kappa(\bar\theta)&=&3\Omega_0\int_0^{\infty}dx\;  y
w(x)\delta^m(\vx)/a.\label{kappa}
\eea
These equations are useful for calculating $\w$.  We obtain
\bea
\w=2\int_0^\infty dx& &\left(\;y^4 S^2(x)\int_0^\infty
dq\; \xi^{gg}\left[\left((x\theta)^2+q^2\right)^{1/2}\right]\right)+\nonumber\\
& &\left(\;9\Omega_0^2(5s-2)^2y^2 w^2(x) a^{-2}\int_0^\infty
dq\; \xi^{mm}\left[\left((x\theta)^2+q^2\right)^{1/2}\right]\right)+\nonumber\\
& &\left(\;6\Omega_0(5s-2)\;y^3 S(x) w(x) a^{-1}\int_0^\infty
dq\; \xi^{gm}\left[\left((x\theta)^2+q^2\right)^{1/2}\right]\right). \label{wx}
\eea
Here $\xi^{gg}$, $\xi^{mm}$, and $\xi^{gm}$ are the galaxy-galaxy,
mass-mass, and galaxy-mass correlations evaluated at the appropriate epoch.
In the case where $s=0.4$, \ie no magnification bias, this reduces to
the ordinary Limber's equation.

We can rephrase the correlation function calculation as an integral in
redshift.  This becomes simply
\bea
\w=2\int_0^\infty dz& &\left(\;H(z)N^2(z)\int_0^\infty
dq\; \xi^{gg}\left[\left((x\theta)^2+q^2\right)^{1/2}\right]\right)+\nonumber\\
& &\left(H^{-1}(z)\;9\Omega_0^2(5s-2)^2y^2 w^2(z)\; (1+z)^2\int_0^\infty
dq\; \xi^{mm}\left[\left((x\theta)^2+q^2\right)^{1/2}\right]\right)+\nonumber\\
& &\left(\;6\Omega_0(5s-2)\;y\; N(z) w(z)\; (1+z)\int_0^\infty
dq\; \xi^{gm}\left[\left((x\theta)^2+q^2\right)^{1/2}\right]\right). \label{wz}
\eea
Here $N(z)=y^2\; S(x) H^{-1}(z)$, $w(z)=w(x)$.  Notice that the
normalisation integral for $N(z)$ still applies,
(Eq. \ref{normS}).
{}From this equation it is possible to calculate $\w$ given a
cosmological model \ie $\Omega_0$, $\Omega_\Lambda$, and a selection
function $N(z)$ if you know, or assume, the three correlation functions.

Let us make life easy and assume that the correlation functions
are all power laws, not necessarily with the same slopes and
evolution.  For each of them assume that in proper coordinates $r$
\be
\xi(r,z)=\left({r \over r_o}\right)^{-\gamma}(1+z)^{-(3+\epsilon)}.
\ee
Here, the exponents $\gamma$ and $\epsilon$ need not be the same for
the correlation functions.  Even more importantly, the proper correlation
lengths $r_0$ need not be the same.  With this assumption
\be
\w\equiv\omega^{gg}(\theta)+\omega^{mm}(\theta)+2\;\omega^{gm}(\theta).
\ee
\bea
\omega^{gg}(\theta)&=& \sqrt{\pi}\;{\Gamma[(\gamma-1)/2] \over
\Gamma[\gamma/2]}\; r_0^\gamma\; \theta^{1-\gamma} \int_0^\infty
dz\;H(z)\;N^2(z)\;x^{1-\gamma}\;(1+z)^{\gamma-3-\epsilon}, \\
\omega^{mm}(\theta)&=& \sqrt{\pi}\;{\Gamma[(\gamma-1)/2] \over
\Gamma[\gamma/2]}\;9\;\Omega_0^2\;(5s-2)^2\; r_0^\gamma\;
\theta^{1-\gamma} \int_0^\infty dz\; H^{-1}(z)\; w^2(z)\;
y^2\;x^{1-\gamma}\;(1+z)^{\gamma-1-\epsilon}, \\
\omega^{gm}(\theta)&=& \sqrt{\pi}\;{\Gamma[(\gamma-1)/2] \over
\Gamma[\gamma/2]}\;3\;\Omega_0\;(5s-2)\; r_0^\gamma\;
\theta^{1-\gamma}\int_0^\infty dz\; N(z)\; w(z)\;y\;
x^{1-\gamma}\;(1+z)^{\gamma-2-\epsilon}.
\eea

The term for the intrinsic clustering $\omega^{gg}$ reduces to the
same result as (BSM), Eqs. (6,7) when taking into account differences
in notation.  Let us make life even easier for ourselves and assume
that $\gamma$ is the same for all three correlation functions.
However allow $\epsilon$ to be different so that
$\epsilon^{mm}=\epsilon+\Delta\epsilon$, and
$\epsilon^{gm}=\epsilon+\Delta\epsilon/2$.  Further assume a linear
bias model so that today $\xi^{gg}(z=0) \equiv b^2\;\xi^{mm}(z=0)$, and
$\xi^{gm}(z=0) \equiv b\;\xi^{gm}(z=0)$.  Now $r_0$ is the correlation length
of the galaxies.  This simplifies the integral for $\w$.
\bea
\w&=&\sqrt{\pi}\;{\Gamma[(\gamma-1)/2] \over
\Gamma[\gamma/2]}\; r_0^\gamma\; \theta^{1-\gamma}\nonumber \\
&\times& \int_0^\infty dz\;H(z)\;
\left[\;N(z)+\;3\;{\Omega_0 \over b}\;(5s-2)\;w(z)\;y\;
(1+z)^{1-\Delta\epsilon/2}\;H^{-1}(z)\right]^2
x^{1-\gamma}\;(1+z)^{\gamma-3-\epsilon}.
\eea
Parametrized this way it is possible to do an analysis of $\w$ in the
same way as BSM while including the effects of the magnification bias.
For a given magnitude limit, assume a cosmological model, \ie
$\Omega_0$, $\Omega_\Lambda$.  Then choose $\gamma$, $\epsilon$,
$\Delta\epsilon$, $b$, and $r_0$.  For a given $N(z)$ this will then
predict the observed $\w$.

As stated in BSM, for $\gamma=1.8$, linear theory predicts that
$\epsilon=0.8$, clustering fixed in comoving coordinates will give
$\epsilon=-1.2$, while clustering fixed in proper coordinates will
have $\epsilon=0$.

If we have further information about $\kappa$ we can improve on our
measurements.  If we can measure image shapes and position angles,
then we can infer the gravitational shear field $p$.  We can then
measure the two point correlation function $\cpp$.  In the weak limit,
$C_{\kappa\kappa}(\theta)=\cpp$ and $\omega^{mm}(\theta)=(5s-2)^2
C_{\kappa\kappa}(\theta)$.  Thus we can remove $\omega^{mm}$ from our
measurements.  This leaves us with the true clustering of the galaxies
plus the cross term galaxy-mass.  If the observational data is of
sufficiently high quality, we can estimate $p$ at any given point on the
sky.  This estimate is of course smoothed over a finite area.  If we
can measure $p$ across the sky, we can infer $\kappa$ through an
inversion procedure such as demonstrated by Kaiser \& Squires (1993),
or Seitz \& Schneider (1995).  With these methods we can obtain an
unbiased estimate of $\kappa$.  There is no problem with
non-linearities since we are certainly in the weak limit and there is
no ambiguity with the mean surface density since it is by assumption zero.
If we know $\kappa$ then we simply subtract the term
$(5s-2)\kappa(\bar\phi)\times N_0(m)$ in Eq. \ref{Nobs}.  We thus
retain an unbiased estimate of the true galaxy clustering.

With this measurement we can thus separate the intrinsic clustering of
galaxies, the clustering of the mass, and the galaxy-mass correlation.
In other words we can separate the galaxy clustering evolution from
the mass clustering evolution, and we can also measure how well light
traces mass.

\section{Discussion}

The standard way of measuring the galaxy number density fluctuations
at intermediate redshifts $z\lessapprox 1$ is through the angular
two-point correlation function $\w $.  Measuring $\w $ is
in principle
straightforward.  You measure the positions and magnitudes of the
galaxies and measure the number of galaxy pairs relative to random.
Small distortions in the telescope optics do not influence the results and
it is not necessary to measure the shape and orientation of the galaxies.
The interpretation of
$\w $ in terms of the three-dimensional correlation function
$\xi(r)$ is complicated through largely unknown luminosity and
density evolution.  Even if $\xi(r)$ were known accurately, this
would only tell us the galaxy distribution, not necessarily the mass
distribution.  The mass distribution would then have to be inferred through a
model dependent biasing scheme, or some other modeling scheme.

Weak lensing by large scale structure is a direct measure of the mass
distribution and it circumvents inferring the mass distribution from
the galaxy distribution.  The problem with weak lensing is that it is
weak.  It requires measuring the shape and orientation of faint
galaxy images.  This is observationally feasible, but quite difficult,
since the images are small and faint, and the possible systematic
errors are the limiting factors.  However, the payoff is immense in
terms of measuring cosmological parameters and the power spectrum of
density fluctuations.

We have presented a new way of measuring the mass density fluctuations
at intermediate redshifts by measuring the angular two-point
correlation function $\w $.  At sufficiently faint magnitudes,
the observed $\w $ will be dominated by weak lensing.  The
method
combines the relative ease at which $\w $ can be determined, with
the relatively simple theoretical interpretation of weak lensing in
terms of cosmological parameters and the statistical properties of the
mass distribution.

At relatively bright magnitudes, $R \lessapprox 23$, weak lensing is
expected to be unimportant.  If the slope of the number counts is
greater/less than $0.4$, the amplitude of $\w $ will decrease
slower/faster than expected from Limber's equation.  In both cases the
amplitude will hit a minimum and then increase with limiting
magnitude.

An equivalent measure is the ratio of number counts in two different
passbands as a function of position in the sky.  This measure can, by
carefully choosing the passbands and magnitudes, be tuned to be nearly
independent of the true clustering of galaxies.  In that case, $\w$ is
a straight measure of weak lensing and the theoretical interpretation
is considerably simpler.

In summary:  The angular two-point correlation function of galaxies is
affected by weak lensing by large scale structure through the
magnification bias.  At faint magnitudes this can be a signicant
effect and must be included in calculations.  For a properly designed
experiment, $\w$ can be used to infer the clustering of the mass, the
clustering of the galaxies, and how well light traces mass.

More work needs to be done.

\bigskip

{\bf ACKNOWLEDGMENTS}

I acknowledge very helpful suggestions and discussions with
T. Broadhurst, T. Brainerd, L. da Costa, W. Freudling, P. Schneider,
and S. White.  MVV290 generously donated time and space.

\bigskip

{\bf REFERENCES}

\medskip

\noindent Bernstein, G.M., Tyson, J.A., Brown, W.R.\ \& Jarvis, J.F.,
1994, ApJ, 426, 516.

\noindent Blandford, R. D., Saust, A. B., Brainerd, T. G., \&
Villumsen, J. V., 1991, MNRAS, 251,600 (BSBV)

\ni Bonnet, H., Mellier, Y., Fort, B., 1994, ApJL, 427, L83

\ni Brainerd, T. G., Smail, I., \& Mould, J. 1995, MNRAS, 275, 781(BSM)

\noindent Broadhurst, T.J., 1995, 5TH Maryland Conference, ``Dark
Matter'', ed. Holt,S.

\noindent Broadhurst, T.J, Taylor, A.N., Peacock, J.A. 1995, ApJ, 438,49

\ni Broadhurst, T.J., Ellis, R.S., \& Glazebrook, K., 1992, Nature, 355,55

\noindent Couch, W.J., Jurcevic, J.S.\ \& Boyle, B.J., 1993, MNRAS,
260, 241

\noindent Efstathiou, G., Bernstein, G., Katz, N., Tyson, J.A.\ \&
Guhathakurta, P., 1991, ApJ, 380, L47

\ni Glazebrook, K., Ellis, R.S., Colless, M., Broadhurst, T.J.,
Allington-Smith, J., Tanvir, N., \& Taylor, K., 1995, MNRAS, 273,157.

\ni Gunn, J.E., 1967, ApJ, 150,737.

\ni Kaiser, N., 1992, ApJ, 388, 272

\ni Kaiser, N., Squires, G., 1993 ApJ, 404, 441

\ni Kaiser, N., Squires, G., Broadhurst, T.J, 1995, ApJ, 449, 460.

\noindent Mould, J., Blandford, R., Villumsen, J., Brainerd, T., Smail, I.,
Small, T., \& Kells, W.,  1994, MNRAS, 271, 31

\noindent Koo, D.C.\ \& Szalay, A.S., 1984, ApJ, 282, 390.

\ni Lilly, S.J., Cowie, L.L., \& Gardner, J.P., 1991, ApJ, 369,79

\ni Miralda-Escud\'{e}, J., 1991, ApJ, 380, 1

\noindent Neuschaefer, L.W., Windhorst, R.A.\ \& Dressler, A.,
1991, ApJ, 382, 32

\noindent Peebles, P.J.E., 1980, `The Large-Scale Structure of the Universe',
Princeton Univ.\ Press.

\noindent Roche, N., Shanks, T., Metcalfe, N.\ \& Fong, R., 1993,
MNRAS, 263, 360

\ni Seitz, C., Schneider, P., 1995, A\&A, 297,287.

\noindent Smail, I., Hogg, D.W, Yan, L., Cohen, J.G., 1995, ApJL {\sl
in press}

\noindent Tyson, A.J., Valdes, F., Wenk, R., ApJL, 349 L19

\noindent Turner, E.L., 1980, ApJL, 242, L135.

\noindent Turner, E.L., Ostriker, J.P., Gott, J.R.III, 1984, ApJ, 284,1.

\noindent Villumsen, J.V. 1995a, MNRAS, {\it in press} V95a.

\noindent Villumsen, J.V. 1995b, MNRAS, {\it submitted}.

\end{document}